\documentclass{aa}  

\usepackage{graphicx}
\usepackage{txfonts}
\begin{document}

   \title{Abundance analysis of red clump stars in  the old, inner disc, open cluster NGC~4337: a twin of NGC~752?\thanks{Based on observations carried out at Paranal Observatory under program 292.D-5043.}}

\author{Giovanni Carraro\thanks{On leave from Dipartimento di Fisica e Astronomia, Universit\'a di Padova, Italy} 
          \inst{1}, Lorenzo Monaco\inst{1}, Sandro Villanova\inst{2}
          }

     \institute{ESO, Alonso de Cordova 3107, 19001,
           Santiago de Chile, Chile\\
              \email{gcarraro,lmonaco@eso.org}
              \and
            Departamento de Astron\'onia, Universidad de Concepci\'on,
               Casilla 169, Concepcion, Chile\\
               \email{svillanova@astro-udec.cl}
             }

\authorrunning{Carraro et al.}
\titlerunning{The old open cluster NGC~4337}

   \date{Received June, 2014; accepted  , 2014}

  \abstract
   {Open star clusters older than $\sim$ 1 Gyr are rare in the inner Galactic disc.  Still, they are objects that hold crucial information for probing the
   chemical evolution of these regions of the Milky Way.}
   {We aim at increasing the number of old open clusters in the inner disc for which high-resolution metal abundances are available.
   Here we report on NGC~4337, which was recently discovered to be an old, inner disc open cluster.}
   {We present the very first high-resolution spectroscopy of  seven clump stars that are all cluster members. We performed a detailed abundance analysis for them.}
   {We find that NGC~4337 is marginally more metal-rich than the Sun, with [Fe/H]=+0.12$\pm$0.05. The abundance ratios of $\alpha$-elements are generally solar. At odds with recent studies on intermediate-age and old open clusters in the Galactic disc, 
    Ba is under-abundant in NGC~4337 compared with the Sun. Our  analysis of the iron-peak elements 
   (Cr and  Ni) does not reveal anything anomalous. Based on these results, we estimate the cluster age to be 1.6$^{+0.1}_{-0.1}$ Gyr,
   and derive a reddening E(B-V)=0.23$\pm$0.05, and an apparent distance modulus $(m-M)_{V}=12.80\pm0.15$. Its distance to the Galactic centre is 7.6 kpc. With this distance and metallicity, NGC~4337 fits the metallicity gradient for the inner Galactic disc fairly well.}
   {The age and metallicity we measured make NGC~4337 a twin of the well-known old open cluster NGC~752. The red clumps of these two clusters bear an amazing resemblance.  But the main sequence of NGC~752 is significantly more depleted in stars than that of NGC~4337. This would mean that
   NGC~752 is in  a much more advanced dynamical stage, being on the verge of dissolving into the  general Galactic field. Our results make
   NGC~4337 an extremely interesting object for further studies of stellar evolution in the critical turn-off mass range 1.1-1.4 $M_{\odot}$.}


   \maketitle
%

\section{Introduction}
Recently, Carraro et al. (2014a) have drawn the attention to the open star cluster NGC~4337, located at  $l =299^{o}.3$, $b=+4^{o}.6$.
It was identified as a particularly relevant
object, being a rare example of an old ($\sim$ 1.5-2.0 Gyr) cluster located inside the solar ring at a Galacto-centric 
distance $R_{GC}$  $\sim$7.5 kpc.
As such, NGC~4337 can help us to improve our knowledge of the inner disc radial abundance gradient and its temporal 
evolution (Magrini et al. 2010, 2014).\\  
This evidence came from the analysis of a deep photometric data-set in the U,B,V, and I passbands, and using the classical
old open IC~4651 (Anthony-Twarog et al. 2009) ridge-line as comparison. 
Previous shallow photoelectric optical photometry had missed the cluster, but more recently, 
it was  recovered by Froebrich et al. (2010) in their search for star clusters  across the Galactic plane using 2MASS.\\
Carraro et al. (2014a) assumed that NGC~4337 shares the same metallicity as IC~4651 from simply 
comparing the distribution of stars in the colour-magnitude diagram (CMD). Based on this, preliminary estimates
of the reddening ($E(B-V) \sim$ 0.26) and the apparent distance modulus ($(m-M)_V \sim$13.00) were derived.\\
With this preliminary fundamental parameter set, NGC~4337 is shown to be an extremely interesting object, sharing the same age and hence turn-off (TO) mass
as IC~4651, NGC~3680, and NGC~752. These text-book clusters have routinely been studied to probe the behaviour of H-burning in the
TO mass range 1.1-1.4$M_{\odot}$, where convective overshooting is expected to take place. Because NGC~4337 is much richer in stars
than these clusters, it is an ideal target to compare its color-magnitude diagrams (CMD) with stellar models (Bertelli et al. 1992; Carraro et al. 1993,1994). The NGC~4337 clump is also particularly rich and well defined, allowing for direct comparisons with stellar models of the He-burning phase in this critical mass range (Girardi et al. 2000).

\begin{table*}
\caption{Clump stars observed with UVES. Coordinates for 2000.0 equinox are reported together with photometry from Carraro et al. (201a4) and heliocentric radial velocities.}
\begin{tabular}{ccrrrrr}
\hline
ID & $RV$ & RA(2000.0) & Dec(2000.0) & V & (B-V) & (V-I) \\
\hline
   &  km/sec & hh:mm:ss.sss & dd:mm:ss.ss & mag & mag \\
\hline
  90 & -17.779$\pm$1.046   &12:24:15.761 & -58:08:50.17  & 13.80  & 1.27 & 1.31\\ 
  91 & -18.180$\pm$1.190   &12:24:39.259 & -58:07:49.69  & 13.83  & 1.31 & 1.40\\ 
  99 & -18.311$\pm$1.026   &12:24:39.482 & -58:13:20.24  & 13.85  & 1.30 & 1.41\\ 
100 & -17.898$\pm$1.227   &12:24:00.948 & -58:05:51.22  & 13.86  & 1.33 & 1.37\\  
102 & -15.976$\pm$1.102   &12:23:27.228 & -58:02:50.24  & 13.86  & 1.35 & 1.45\\  
115 & -18.224$\pm$1.102   &12:23:56.662 & -58:07:25.21  & 13.93  & 1.32 & 1.36 \\
128 & -17.920$\pm$1.127   &12:24:44.945 & -58:02:09.96  & 14.06  & 1.33 & 1.44\\ 
\hline
\end{tabular}
\end{table*}

\noindent
In this paper we follow the study of Carraro et al. (2014a) up and present for the first time high-resolution spectroscopy obtained with the FLAMES-UVES spectrograph at ESO Paranal. These data are used to measure the abundance of several elements 
(such as  $\alpha$ , neutron-capture, and iron peak) for the stars in the red clump of NGC~4337. 
The knowledge of the metal abundance, in turn, allows us to provide more solid estimates of the cluster fundamental parameters (mostly age and distance),  and discuss its properties in the framework of the Milky Way structure and chemical evolution.\\

\noindent
The paper is organised as follows: In Sect.~2 we present the observational material and describe how radial velocities are derived.
Section~3 is devoted to the abundance analysis, while in Section~4 we discuss the results of our analysis in detail.
Section~5 summarises our findings.

\section{Observations and data reduction}
Observations were taken in service mode on the night of March
30, 2014 using the multi-object fibre-fed FLAMES facility mounted at the
ESO-VLT/UT2 telescope at the Paranal Observatory (Chile).  Two 2775s exposures
were taken simultaneously with the  GIRAFFE medium-resolution spectrograph, and
the red arm of the UVES high-resolution spectrograph. GIRAFFE was configured in
Medusa mode in the setup HR15N, which covers the wavelength range 
6470--6790\AA\, at a resolution of $R$=17,000. This set-up was chosen because it  simultaneously includes
the Li resonance doublet and the H$_\alpha$line.
The analysis of this data-set will be presented in a forthcoming paper.\\
The UVES spectrograph was, instead,
set up around a 5800\AA\, central wavelength, thus covering the 4760--6840\AA\, wavelength
range and providing a resolution of R$\simeq$47,000.\\

\noindent
Here we discuss the UVES data.
The UVES fivers were placed on probable 
red-clump targets, and are shown on  top of the cluster's  $V\, vs.~ V-I$
colour-magnitude diagram in Fig.~1.  
Table\,1  presents the target star IDs,
coordinates, and $B$, $V$, and $I$ photometry from  Carraro et al. (2014a).
The data were reduced using the ESO CPL based FLAMES-UVES pipeline version
5.3.0\footnote{\url{http://www.eso.org/sci/software/pipelines/}} to
extrac the individual fibre spectra. \\

\noindent
The spectra were eventually normalised using the standard IRAF task {\tt
continuum}. Radial velocities were computed using the IRAF/{\tt fxcor} task to
cross-correlate the observed spectra with a synthetic spectrum from the library of
Coelho et al. (2005)  with stellar parameters  $T_{\rm eff}$=4750\,K, log\,$g$=2.0, solar
metallicity, and no $\alpha$-enhancement. The IRAF {\tt rvcorrect} task was used
to calculate the correction from geocentric velocities to heliocentric.\\

\noindent
The seven stars observed with UVES have the same heliocentric radial velocity within the uncertainty (see Table~1),
supporting the idea that they are all cluster members. Therefore,
the mean heliocentric radial velocity of NGC~4337  RV = -17.75$\pm$0.81 km/sec.

   \begin{figure}
   \centering
   \includegraphics[width=\columnwidth]{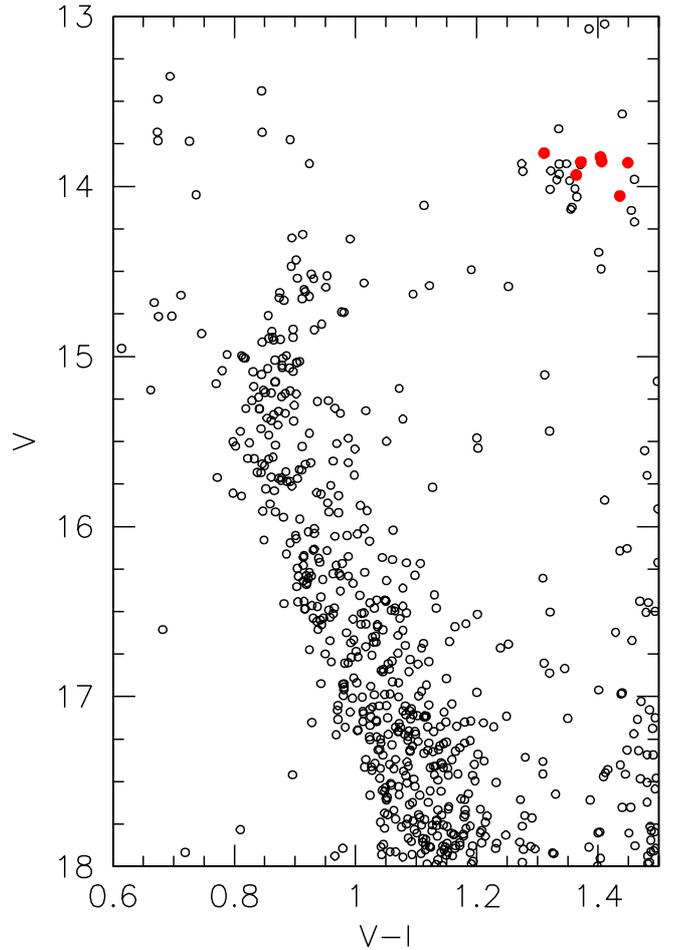}
   \caption{Color-magnitude diagram of NGC~4337. Red filled circles identify the seven red clump stars observed with UVES.}%
    \end{figure}

\begin{table}
\caption{Iron abundance and photospheric parameters for clump stars in NGC~4337.}
\begin{tabular}{crrrr}
ID     &        $T_{\rm eff}$   & log(g) &  [Fe/H]   & $v_{t}$\\
\hline
         &   $^{o}K$ & & & km/sec \\
\hline
 90   & 4790  & 2.65  & +0.10  & 1.19\\
 91   & 4730  & 2.70  & +0.12  & 1.21\\
 99   & 4730  & 2.70  & +0.13  & 1.19\\
100  & 4760  & 2.60  & +0.09  & 1.27\\
102  & 4720  & 2.60  & +0.11  & 1.20\\
115  &  4760 & 2.70  & +0.14  & 1.17\\
128  & 4720  & 2.55  & +0.13  & 1.13\\
\hline
\end{tabular}
\end{table}

\section{Abundance analysis}
The chemical abundances for Na, Mg, Al, Si, Ca, Ti, Cr, Fe, and Ni were
obtained using the equivalent widths of species transition lines (EW  method, as detailed in 
Marino et al. (2008)
For C, N, O, Y, Ba, La, and Eu, whose lines are affected by blending, we used the
spectrum-synthesis method. For this
purpose we calculated five synthetic spectra with different abundances for the
elements, and estimated the best-fitting value as the one that minimises the
$r.m.s.$ scatter. Only lines un-contaminated by telluric lines were used.
ATLAS9 (Kurucz 1970) model atmospheres were used for the
EW and spectrum-synthesis methods. 
The initial atmospheric parameters for the model atmosphere 
were assumed to be  those typical for an RGB star of an open cluster, that is 
$T_{\rm eff}$=4500 \,K, log\,$g$=2.5, $v_{\rm t}$=1.20 km/s, and [Fe/H]=0.0.
We then  refined them during the abundance analysis. 
As a first step, atmospheric models were calculated using ATLAS9 (Kurucz 1970)
with the initial estimates of $T_{\rm eff}, \log g$,
 $v_{\rm t}$, and  [Fe/H].
 
The values of T$_{\rm eff}$, v$_{\rm t}$, and $\log g$ were adjusted and new
atmospheric models calculated in an interactive way to remove trends in
excitation potential (EP) and equivalent width $vs.$ abundance for
T$_{\rm eff}$ and v$_{\rm t}$, respectively, and to satisfy the ionisation
equilibrium for $\log g$. \ion{Fe}{I} and \ion{Fe}{II}  were used for this purpose.
The [Fe/H] value of the model was changed at each iteration according to the
output of the abundance analysis (see Table~2).
The local thermodynamic equilibrium (LTE) program MOOG (Sneden 1973) was used
for the abundance analysis.\\

\noindent
In all the analysis, we closely followed Carraro et al. (2014b). We therefore refer to this study
for additional details.
We briefly recall that typical internal errors are $\Delta(T_{\rm eff})$=50 K,  $\Delta\log g$=0.2 dex, $\Delta v_{\rm t}$=0.10 km/s, 
and $\Delta$[Fe/H]=0.05 dex.

\begin{table*}
\caption{Individual abundances for NGC~4337 members. In addition to  iron, presented as [Fe/H],  abundances are provided as [X/Fe], where X is the chemical species. With $\alpha$ we indicate the average of the $\alpha$-elements O, Mg, Si, Ca, and Ti abundances.}
\begin{tabular}{crrrrrrrrrrrr}
\hline\hline
       ID   &   C  &  N  &   O  &  CNO &  Na &  Mg &  Al & Si  & Ca   &   Ti & $\alpha$    \\ 
\hline
 90 &-0.14 &0.26 &-0.03 &-0.02 &0.33 &0.10 &0.12 &0.13 &-0.02 &-0.12 & 0.00\\  
 91 &-0.16 &0.33 &-0.01 &-0.00 &0.32 &9.99 &0.15 &0.13 &-0.05 &-0.13 &-0.02\\ 
 99 &-0.18 &0.35 & 0.02 & 0.01 &0.39 &0.01 &0.12 &0.09 & 0.00 &-0.06 & 0.01\\ 
100 &-0.12 &0.33 &-0.07 &-0.03 &0.34 &0.05 &0.20 &0.07 &-0.04 &-0.12 &-0.03\\ 
102 &-0.14 &0.32 &-0.05 &-0.02 &0.36 &0.10 &0.18 &0.06 &-0.05 &-0.07 &-0.02\\ 
115 &-0.15 &0.27 &-0.06 &-0.04 &0.33 &0.05 &0.19 &0.07 &-0.02 &-0.06 & 0.00\\ 
128 &-0.19 &0.29 &-0.12 &-0.08 &0.34 &0.04 &0.17 &0.07 &b0.00 &-0.16 &-0.03\\ 
\hline\hline
       ID   &   V &  Cr  &  Fe &  Co &  Ni & Cu  &  Zn  &   Y  &  Zr  & Ba   &  Ce  &  Eu \\
\hline
 90 &0.58 &-0.02 &0.10 &0.24 &0.09 &0.12 & 9.99 &-0.26 & 9.99 &-0.06 &-0.25 & 0.08\\
 91 &0.59 &-0.01 &0.12 &0.25 &0.11 &0.03 & 9.99 &-0.18 &-0.33 &-0.09 &-0.23 & 0.14\\
 99 &0.52 &-0.06 &0.13 &0.18 &0.11 &0.15 &-0.36 &-0.23 & 9.99 &-0.02 &-0.38 & 0.16\\
100 &0.60 & 0.02 &0.09 &0.28 &0.10 &0.09 &-0.38 &-0.23 &-0.30 &-0.16 & 9.99 & 0.09\\
102 &9.99 &-0.05 &0.11 &0.25 &0.12 &0.10 &-0.49 &-0.21 &-0.39 &-0.07 &-0.20 & 0.05\\
115 &0.58 &-0.02 &0.14 &0.26 &0.13 &0.16 &-0.35 &-0.03 &-0.49 &-0.04 &-0.23 & 0.12\\
128 &0.48 &-0.04 &0.13 &0.29 &0.14 &0.10 & 9.99 &-0.33 & 9.99 &-0.15 &-0.25 &-0.01\\
\hline
\end{tabular}
\end{table*}

The line lists for the chemical analysis were obtained from many sources 
(Gratton et al. 2003, VALD \& NIST\footnote{\url{http://physics.nist.gov/PhysRefData/ASD/}}; 
McWilliam \& Rich 1994; McWilliam 1998, SPECTRUM\footnote{\url{http://www.phys.appstate.edu/spectrum/spectrum.html}}, 
and SCAN\footnote{\url{http://www.astro.ku.dk/~uffegj/}}), 
and the log\,gf were calibrated using the solar-inverse technique and by the spectral synthesis method, as
discussed in full detail by Villanova et al. (2009). 
For this purpose we used the high-resolution, high S/N NOAO solar spectrum (Kurucz et al. 1984). 
The solar abundances we obtained with our line list are reported in Table~4,
together with those given by Grevesse \& Sauval (1998) for comparison. 
We emphasise that all the line-lists were calibrated on the Sun, including those used for the spectral synthesis.
We provided the line list  as a long table in Carraro et al. (2014b).
In addition, the  C content was obtained from the C$_2$ system at 563.2 nm, and N from the CN lines
at 634 nm.\\

\noindent
Abundances for C, N, and O were determined all together in an
interactive way to take into account any possible molecular coupling
of these three elements.
Our targets are objects evolved off the main sequence, so some evolutionary
mixing is expected. This can affect the primordial C and N abundances separately,
but not the total C+N+O content because these elements are transformed one
into the other during the CNO cycle.\\

 \begin{figure}
 \centering
   \includegraphics[width=\columnwidth]{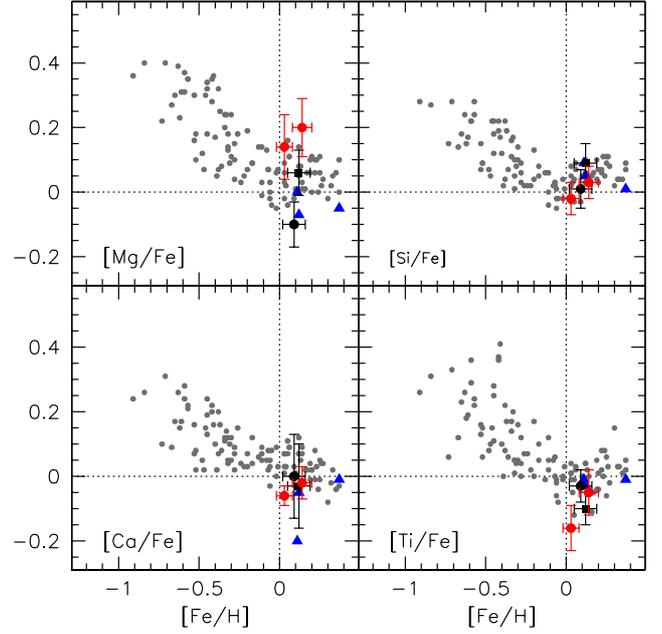}
   \caption{Comparison between NGC~4337 mean $\alpha-$element ratios (black solid square)  including error bars derived from Table~5, with data 
from old open clusters in the inner disc from Magrini et al. (2010; blue triangles) and inner disc giants from Bensby et al. (2010; grey dots). The black solid circle indicates Trumpler~20 from Carraro et al. (2014b), and the red circles are NGC~4815 and NGC~6705 from Magrini et al. (2014).}
  \end{figure}

\section{Results of abundance analysis}
In this section, we discuss in detail the outcome of the abundance analysis and its impact
on the determination of the cluster parameters.

\subsection{Metallicity and $\alpha$ elements}
Table~2 lists the metallicity, measured as [Fe/H], of the seven clump stars we observed with UVES. The mean value is 
[Fe/H] = 0.12$\pm$0.01, with very small dispersion (see also Table~4). This is the very first estimate of the  metallicity  of NGC~4337 , and the value we obtained is well within the expectations for a cluster located inside the solar ring (Magrini et al 2010,2014). It is indeed quite similar to Trumpler~20 (Carraro et al. 2014b, Magrini et al. 2014), an open cluster located very close to NGC~4337.\\

\noindent
In Fig.~2 we plot the trend of averaged  $\alpha-$element abundance of NGC~4337 versus metallicity, an compare them with different objects, 
 The comparison is against K giants in the inner disc taken from Bensby et al. (2010) study. 
 Besides, we also show  a few inner disc open clusters: Trumpler 20 (Carraro et al. 2014b), NGC~4815 and NGC~6705 (Magrini et al. 2014), and NGC~6404, NGC~6503, and NGC 6192 (Magrini et al. 2010). Solar abundances are indicated with dotted lines.\\
The agreement is in general very good, and all these clusters appear to be members of the same stellar population. The exception is [Mg/Fe]
which shows  a quite significant scatter. This was also reported present in Magrini et al. (2014), because Trumpler~20 shows a Mg under-abundance with respect to the Sun, similar to the clusters studied in Magrini et al. (2010). NGC~4337, instead, comfortably lies in the general trend defined by K giants. This is true not only for Mg, but for all the other $\alpha-$ elements.

\subsection{Iron peak elements}
We culled from the literature the  abundances of thin- and thick-disc stars iron peak element (Ni and Cr, (Reddy et al. 2003, 2006)
and compare them with our results for NGC~4337 in Fig.~3. We also considered, as in Fig~2, old open clusters in the inner disc.
The upper panel shows Cr abundance ratio, while the lower show Ni. Solar abundances are indicated with dotted lines.\\
As emphasised already in the past (Carraro et al. 2014b), the sample of Reddy et al. do not have many stars
with the same metallicity as inner disc star clusters. In spite of this limitation, clusters and stars shows essentially the same trend.
NGC~4337, however, shows a significant overabundance in [Ni/Fe], compared to other clusters. In [Cr/Fe],
lastly, it shows a basically solar abundance.

\begin{figure}
\centering
\includegraphics[width=\columnwidth]{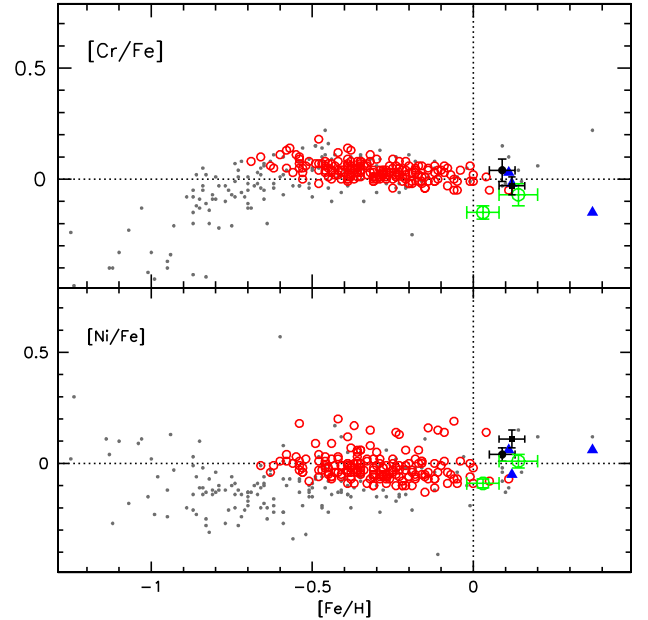}
\caption{Mean Cr (lower panel),  and Ni (upper panel) abundance ratios for NGC~4337 giants (filled black squares), compared with thin-disc stars (large  red open circles),
thick-disc stars (grey symbols), and inner disc open clusters (Magrini et al. 2010, blue triangles), Magrini et al. (2014, green empty circles ),
and Trumpler~20 (black filled circle).}
\end{figure}

\subsection{Neutron-capture elements}
There currently is a lively discussion on the Ba and the Y overabundance in star clusters and associations and its origin.
According to recent studies (Mishenina et al. 2013), star clusters for which Ba and Y were measured tend to exhibit a Ba overabundance with respect to the Sun, and this overabundance correlates with age, namely it increases at decreasing age. In addition, for clusters younger than 3-4 Gyr, the Ba abundance shows a large scatter.\\
Our results are shown in Fig.~4, where NGC~4337  is compared with Trumpler~20 and the sample
of old open clusters from Mishenina et al. (2013). Thin- and thick-disc stars from Reddy et al. (2003,2006) are also plotted.
Both NGC~4337 and Trumpler~20 are underabundantin Y, and  are the only open clusters known to shown such underabundance
above solar metallicity. 
NGC~4337 appears to be  Ba underabundant with respect to the Sun. This behaviour is unique among open clusters more metal-rich than the Sun.

\begin{figure}
\centering
\includegraphics[width=\columnwidth]{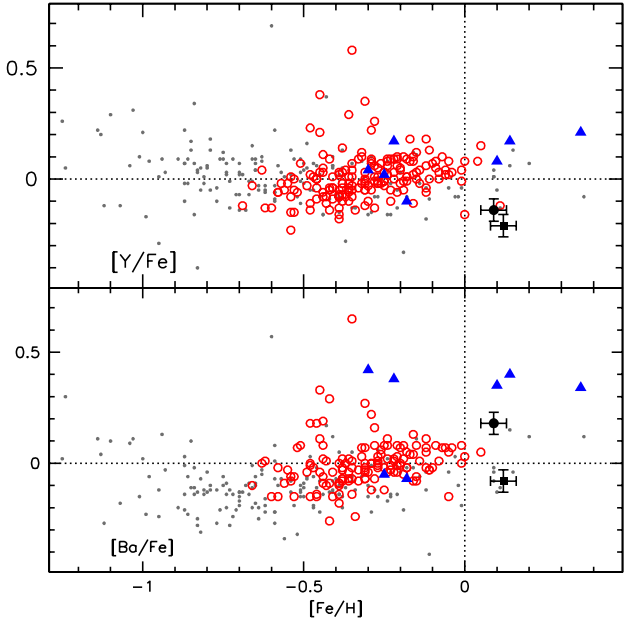}
\caption{Mean Ba (lower panel) and Y (upper panel) abundance ratios for NGC~4337 giants (filled black squares), compared with thin-disc stars (large red open circles),
thick-disc stars (grey symbols), and  old open clusters (bluetriangles, from Mishenina et al. 2013)). Lastly, the filled black circle indicate Trumpler~20. }
\end{figure}

\begin{table}
\caption{Mean NGC~4337 abundance ratios.}
\begin{tabular}{crr}
\hline\hline
Abundance ratio & Mean~~~ & $\sigma$~~~~~~ \\
\hline                
<[C/Fe]>      & -0.15$\pm$0.04  &0.02$\pm$0.01\\
<[N/Fe]>      & 0.31$\pm$0.02   &0.03$\pm$0.01\\
<[O/Fe]>      & -0.05$\pm$0.02  &0.05$\pm$0.01\\
<[CNO/Fe]> & 0.04$\pm$0.03  &0.03$\pm$0.01\\
<[Na/Fe]>    & 0.34$\pm$0.03         &0.02$\pm$0.01\\
<[Mg/Fe]>    & 0.06$\pm$0.04        &0.04$\pm$0.01\\
<[Al/Fe]>     & 0.16$\pm$0.02          &0.03$\pm$0.01\\
<[Si/Fe]>     & 0.09$\pm$0.03          &0.03$\pm$0.01\\
<[Ca/Fe]>    & -0.03$\pm$0.04       &0.02$\pm$0.01\\
<[Ti/Fe]>      & -0.10$\pm$0.03         &0.04$\pm$0.01\\
<[alpha/Fe]> & -0.01$\pm$0.03   &0.01$\pm$0.00\\
<[V/Fe]>       & 0.56$\pm$0.07           &0.05$\pm$0.01\\
<[Cr/Fe]>     & -0.03$\pm$0.03        &0.03$\pm$0.01\\
<[Fe/H]>     & 0.12$\pm$0.05          &0.02$\pm$0.00\\
<[Co/Fe]>  & 0.25$\pm$0.03        &0.04$\pm$0.01\\
<[Ni/Fe]>    & 0.11$\pm$0.03         &0.02$\pm$0.00\\
<[Cu/Fe]>  & 0.11$\pm$0.02        &0.04$\pm$0.01\\
<[Zn/Fe]> & -0.40$\pm$0.03       &0.06$\pm$0.02\\
<[Y/Fe]> & -0.21$\pm$0.03        &0.09$\pm$0.02\\
<[Zr/Fe]> & -0.38$\pm$0.04       &0.08$\pm$0.03\\
<[Ba/Fe]> & -0.08$\pm$0.02      &0.05$\pm$0.01\\
<[Ce/Fe]> & -0.26$\pm$0.03      &0.06$\pm$0.02\\
<[Eu/Fe]> & 0.09$\pm$0.02       &0.06$\pm$0.02\\
\hline
\hline
\end{tabular}
\end{table}

\section{Fundamental parameters of NGC~4337}
We derive in this section updated estimates of the basic parameters of NGC~4337 based on the results of the previous section.
To this aim we generate isochrones using the stellar models from the Padova group (Bressan et al. 2012) by adopting a metal content 
[Fe/H=0.12$\pm$0.05] which  translates into Z=0.025. We generated several isochrones, and obtained
the best fit for an age of 1.6$\pm$0.1 Gyr. The quality of the fit was judged by visually inspecting the CMD, with particular attention
on the main sequence (MS) turn-off (TO) region, the MS slope, and the red clump. The best fit is shown in Fig.~5.\\
As a result, we estimate the reddening towards NGC~4337 to be E(B-V)=0.23$\pm$0.05, and its apparent distance modulus
(m-M)$_{V}$=12.80$\pm$0.15. This implies a heliocentric distance of 2.6$\pm$0.2 kpc, which places the cluster at 200 pc above the Galactic plane and 7.6 kpc from the Galactic centre. \\

\noindent
Having a more solid handle on the cluster metallicity and reddening, we re-compute - as an internal sanity check- the stars' T$_{\rm eff}$ using Alonso et al. (1999) calibrations for (B-V) colours.  
We find that the differences in T$_{\rm eff}$  do not exceed 80~$^{o}K$, and are in most case smaller than 20~$^{o}K$. 
This is because the reddening adopted to derive T$_{\rm eff}$  estimates in Table~2 is only 0.03
mag larger (0.26, Carraro et al. 2014a)  than  the reddening estimated in this paper, and the metallicity only marginally super-solar.\\

\begin{figure}
\centering
\includegraphics[width=\columnwidth]{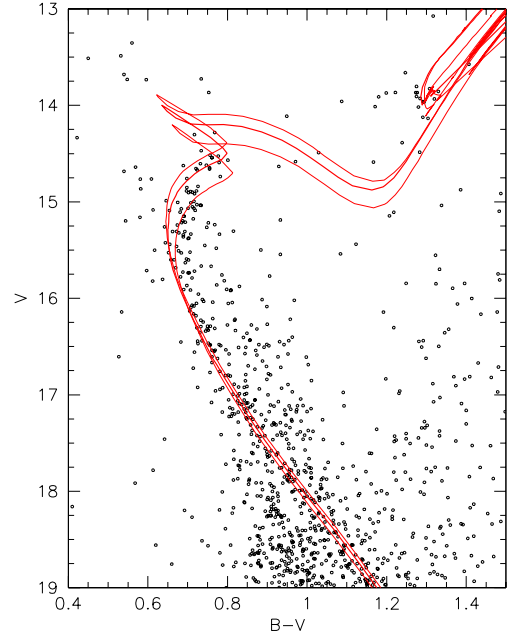}
\caption{V/B-V CMD of NGC~4337 stars within 5 arcmin from  the cluster centre. Super-imposed (red line) are Z=0.025 metallicity isochrones
for the ages of 1.5, 1.6, and 1.7  Gyr. The inferred reddening and apparent distance modulus are E(B-V)=0.23$\pm$0.05 and (m-M)$_{V}$=12.80$\pm$0.15}
\end{figure}

\begin{figure}
\centering
\includegraphics[width=\columnwidth]{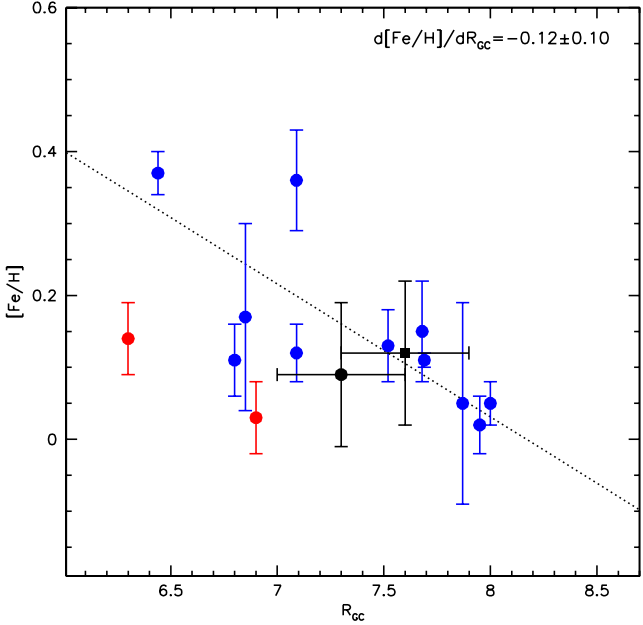}
\caption{The inner disc  radial abundance gradient from Magrini et al. (2010, 2014; blue and red symbols, respectively) with the addition of Trumpler~20 (black circle) and NGC~4337(black square). The dotted line is a weighted least square fit to the points, that yields $-0.12$ as value of the slope.}
\end{figure}

\noindent
In Fig.~6 we show an updated version of the radial metallicity gradient in the inner disc. Differently to Carraro et al. (2014b) we added 
NGC~4337 (solid square) and NGC~6705 and NGC~4815 from Magrini et al. (2014, red symbols). NGC~4337 follows
the trend defined by Magrini et al. (2010) sample,  and Trumpler~20 very well.
Including of NGC~4815 and NGC~6705, which clearly seem do deviate from the main relation, significantly decreases the slope
of the gradient to d[Fe/H]/dR = $-0.12\pm0.10$ dex/kpc, and increases the metallicity scatter at any distance. We caution, however, that these
two clusters are significantly younger. Lastly, we note that an almost flat gradient would result if the two most metal-rich clusters, namely NGC~6583 and NGC~6253, were removed.

\section{Discussion and conclusions}
We have presented the first high-resolution spectroscopic study of the inner disc, old open cluster NGC~4337. Our main results are
as follows:

\begin{description}
\item ({\it i}) NGC~4337 has a marginally super-solar metal content ([Fe/H]=0.12$\pm$0.05)
\item - the $\alpha-$ elements are essentially solar, while we detect  a significant underabundance of Ba, unexpected in a star cluster 
with this metallicity. Y is also underabundant, similarly to Trumpler~20, and at odds with other clusters;
\item ({\it ii}) we refine the results of Carraro et al. (2014a),  and estimate an age of 1.6$\pm$0.1 Gyr for the cluster.
\item ({\it iii}) with this metallicity, and a Galacto-centric distance of 7.6 kpc, NGC~4337 fits  in the inner disc radial metallicity gradient very well. 
\end{description}

\noindent
Carraro et al. (2014a)  compared NGC~4337 with IC~4651, a well-known old open cluster, and suggested that NGC~4337 is slightly younger than IC~4651, basing on the CMD morphology and the assumption of identical metallicity. The comparison was 
very qualitative, since it was based on an averaged IC~4651 ridge-line, which is easy to draw for the MS, but can be much more complicated
for the clump, which, for this age range, is typically quite broad, both in colour and in magnitude.\\

\noindent
Now that we independently estimate the age and metallicity of NGC~4337, we can compare it again with IC~4651. According to Anthony-Twarog et al. 2009), IC~4651 is 1.5 Gyr old for a metallicity [Fe/H]=0.12 (see also Carretta et al. 2004), identical to  the metallicity of NGC~4337.  It has a reddening E(B-V)$\sim$0.12, and an apparent distance modulus (m-M) $\sim$10.40. 

\begin{figure}
\centering
\includegraphics[width=\columnwidth]{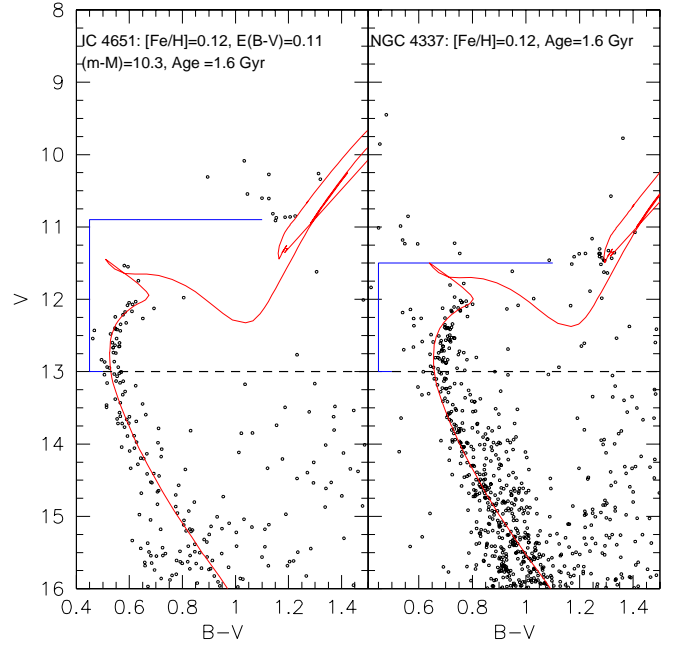}
\caption{Comparison of NGC~4337 (right panel) with IC~4651 (left panel). The dashed line indicates the TO magnitude of IC~4651, which NGC~4337 has been shifted to for comparison purposes. The blue segment shows the estimate of the parameter $\Delta$V. Best-fit values for the clusters' fundamental parameters are written in the top of the panels.}
\end{figure}

\noindent
We compare the two clusters in Fig.~7.  The blue segment shows the estimate of the parameter $\Delta$V, defined as the magnitude difference between the blue TO and the red clump magnitude. This parameter is a very well-known relative age indicator (Carraro \& Chiosi 1994), independent on reddening, and with some dependence on metallicity.
A higher $\Delta$V values means an older age. In this case, the $\Delta$V for IC~4651 and NGC~4337 are 2.1 and 1.6 mag.\\
Super-imposed on the  NGC~4337 stars is the best-fit isochrone discussed in Sect.~5. Since NGC~4337 and IC~4651 have the very same metallicity, we super-imposed the same isochrone on top of IC~4651. A 1.6 Gyr isochrone misses the clump, which is clearly brighter  and at the same time does not reproduce the TO curvature and extension well.\\
This evidence, in tandem with the higher value of $\Delta$V, suggests that IC~4651 is significantly older than NGC~4337. In our scale (the Padova models), IC~4651 would in fact be about 1.9 Gyrs old. This age would be consistent with a lower  reddening E(B-V) $\sim$ 0.09 mag, and a smaller  apparent distance modulus  (m-M)$_V$ $\sim$10.15 mag.\\
Therefore, although they share the same metallicity,  NGC~4337 and IC~4651 do have significantly different ages.\\

\noindent
We found NGC~752 (Anthony-Twarog \& Twarog 2006)  to be a stronger case for a NGC~4337 twin. NGC~752 has a marginally lower metallicity ([Fe/H]=0.08$\pm$0.04, Carrera \& Pancino 2011) which appears to be very similar to NGC~4337 one within the uncertainties however.\\
The comparison is illustrated in Fig.~8, where the CMD of NGC~4337 is compared to the CMD of NGC~752 from Francic (1989).
We adopted for both cluster the same metallicity ([Fe/H=0.012) and the same age (1.6 Gyr). The fit is remarkably good in both cases, strongly suggesting that these two clusters share the same chemical and evolutionary phase.\\

\noindent
While the clump in both clusters is equally populated and possibly delineates the same evolutionary phase (Girardi et al. 2000), the real difference among these two clusters is in the MS, which for  NGC~752 shows a significant star-smearing immediately below the TO. It is indeed impossible to follow the cluster MS below  V $\sim$ 11.50 mag.
This is interpreted as a signature that NGC~752 has undergone quite a strong dynamical evolution and is on the verge of dissolving
into the general Galactic field. Since the two clusters are coeval, this suggests that either NGC~4337 formed as a more massive cluster than NGC~752, or that the latter experienced a much stronger tidal  interaction with the Milky Way than the former.\\

\noindent
Because it is more massive, and hence has retained more stars, NGC~4337 is an ideal object in which to compare the star distribution in the CMD against stellar evolution models in a statistical manner.

 \begin{figure}
\centering
\includegraphics[width=\columnwidth]{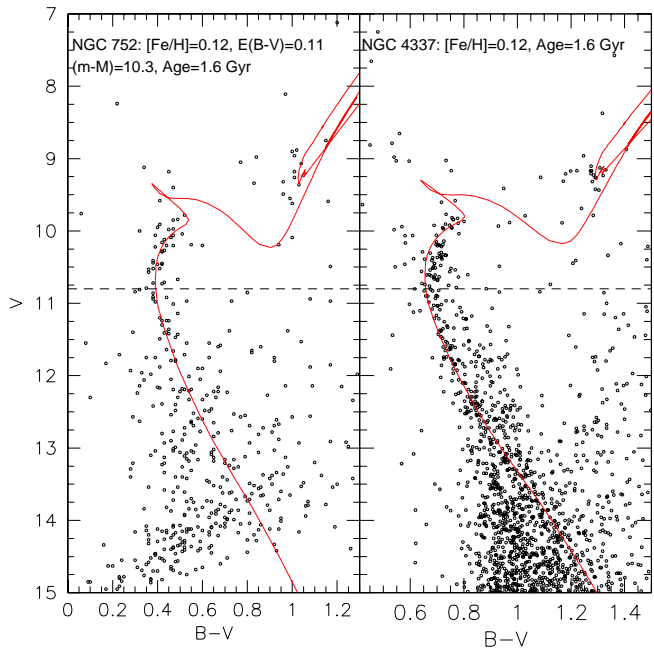}
\caption{Comparison of NGC~4337 (right panel) with NGC 752 (left panel). The dashed line indicates the TO magnitude of NGC~752, which NGC~4337 has been shifted to for comparison purposes.  Best-fit values for the clusters' fundamental parameters are written in the top of the panels .}
\end{figure}

\begin{acknowledgements}
S. Villanova gratefully acknowledges the support provided by FONDECYT program N. 1130721. G. Carraro expresses his gratitude to B. Twarog  for his very useful comments. The ESO Director General is deeply thanked for granting time to this project via the Director General Discretionary Time  (DDT).
\end{acknowledgements}


\end{document}